\documentclass{moriond}

\def\be{\begin{equation}}
\def\ee{\end{equation}}
\def\bea{\begin{eqnarray}}
\def\eea{\end{eqnarray}}

\usepackage{lineno}
\usepackage{xspace}
\newcommand*{\xs}{cross-section\xspace}
\newcommand*{\xss}{cross-sections\xspace}
\usepackage{amsmath}
\usepackage{booktabs}
\newcommand*{\ifb}{\ensuremath{\text{fb}^{-1}}\xspace}
\newcommand*{\fb}{\ensuremath{\text{fb}}\xspace}
\newcommand*{\pp}{proton-proton\xspace}
\newcommand*{\ggH}{\ensuremath{ggH}\xspace}
\newcommand*{\VBF}{\ensuremath{VBF}\xspace}
\newcommand*{\VH}{\ensuremath{VH}\xspace}
\newcommand*{\ttH}{\ensuremath{t\bar{t}H}\xspace}
\newcommand*{\Hbb}{\ensuremath{H \rightarrow b\bar{b}}\xspace}

\newcommand*{\HWW}{\ensuremath{H \rightarrow WW^*}\xspace}
\newcommand*{\GeV}{\ensuremath{\text{Ge\kern -0.1em V}}\xspace}
\newcommand*{\TeV}{\ensuremath{\text{Te\kern -0.1em V}}\xspace}

\newcommand*{\ptH}{\ensuremath{p_{\text{T}}^H}\xspace}
\newcommand*{\ptV}{\ensuremath{p_{\text{T}}^V}\xspace}

\usepackage[sorting=none,
            style=numeric-comp,
            giveninits=true]{biblatex}
\bibliography{my_bibliography}

\DeclareFieldFormat{paperlink}{%
 \iffieldundef{url}{#1}%
 {\href{\thefield{url}}{#1}}
}

\DeclareDelimFormat*{labelnamepunct}{\addcomma\space}
\DeclareFieldFormat*{title}{\textit{#1}\addcomma\space}
\DeclareFieldFormat*{volume}{\textbf{#1}}
\DeclareFieldFormat*{url}{\url{#1}}

\renewbibmacro*{journal+issuetitle}{%
  \printfield[noformat]{journaltitle}%
  \addspace
  \printfield{volume}%
  \addspace
  \parentext{\printfield[noformat]{year}}%
  \iffieldundef{pages}{}{\addspace\printfield{pages}}%
}

\DeclareBibliographyDriver{article}{%
  \usebibmacro{begentry}%
  \usebibmacro{author}%
  \printdelim{labelnamepunct}%
  \usebibmacro{title}%
  \printtext[paperlink]{%
    \usebibmacro{journal+issuetitle}%
  }%
  \addperiod\addspace
  \usebibmacro{finentry}
}

\DeclareBibliographyDriver{report}{%
  \usebibmacro{begentry}%
  \usebibmacro{author}%
  \printdelim{labelnamepunct}%
  \usebibmacro{title}%
  \printdelim{labelnamepunct}%
  \printfield{number}%
  \printdelim{labelnamepunct}%
  \printtext[paperlink]{%
    \printfield{url}%
  }%
  \addperiod\addspace
  \usebibmacro{finentry}
}

\newcommand*{\fig}[4][0.6]{ % #1 = page_width_multiplier, #2 = filename, #3 = caption, #4 = label
	\begin{figure}
		\centering
		\includegraphics[width=#1\textwidth]{#2}
		\caption{#3}
		\label{#4}
	\end{figure}
 }

\newcommand{\Photo}{\includegraphics[height=35mm]{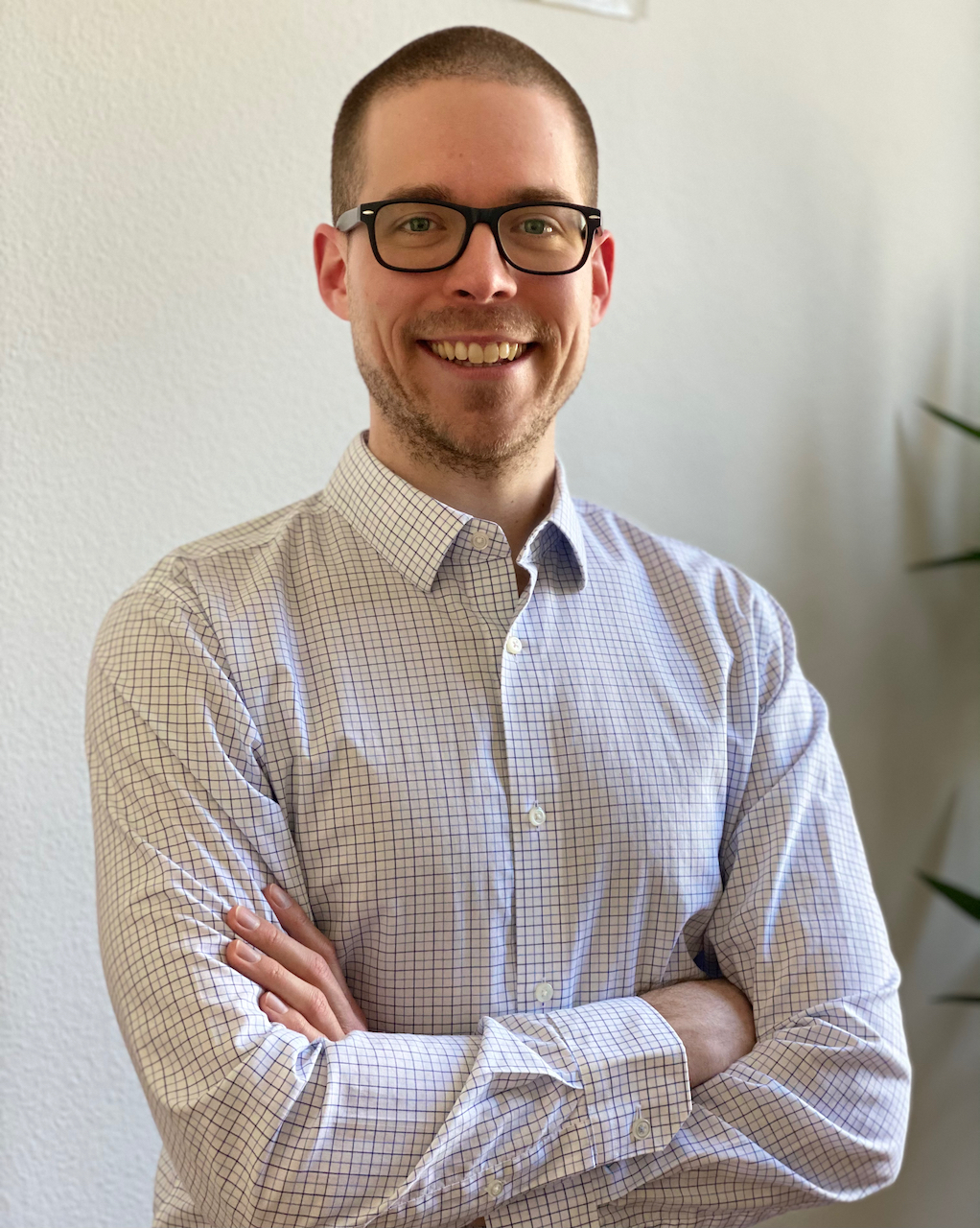}}

\begin{document}
% \linenumbers
\vspace*{4cm}
\title{Higgs cross-section and properties at ATLAS and CMS}

\author{Sébastien Rettie, on behalf of the ATLAS and CMS Collaborations}

\address{Department of Physics \& Astronomy, University College London,\\
Gower St, Bloomsbury, London WC1E 6BT, UK}

\maketitle\abstracts{
Recent measurements of Higgs boson \xs and properties are presented using up to 139~\ifb of \pp collision data delivered by the Large Hadron Collider at $\sqrt{s}~=~13~\TeV$ and recorded by the ATLAS and CMS detectors.
Three measurements are discussed.
The first is the measurement of Higgs boson production with sizeable transverse momentum decaying to a $b\bar{b}$ pair.
The remaining two measurements exploit the \HWW decay channel in various production modes: gluon fusion, vector boson fusion, and production in association with a $W$ or $Z$ boson.
The results presented are compatible with Standard Model predictions.
}

\section{Introduction}

The dataset collected at the Large Hadron Collider (LHC)~\cite{LHC} during 2015-2018 (Run 2) provides an exquisite opportunity to explore the rich phenomenology of the Higgs boson, and to test the validity of the Standard Model (SM) of particle physics.
So far, measurements carried out by the ATLAS~\cite{ATLAS} and CMS~\cite{CMS} experiments are consistent with the SM Higgs boson.
The next logical steps are to increase the precision of these measurements, and carry out differential measurements.
In particular, the simplified template \xs (STXS)~\cite{STXS} framework allows for the categorization of events into Higgs production modes by focusing on key quantities such as the transverse momentum of the Higgs boson (\ptH) and the number of jets in the final state.
The main production modes are gluon fusion (\ggH), vector boson fusion (\VBF), associated production with a $W$ or $Z$ boson (\VH, $V=W,Z$), and associated production with top quarks (\ttH)~\cite{LHCHWG3}.
In this article, the most recent results by both experiments are presented: Section~\ref{sec:Hbb} describes the analysis of \Hbb decays in the boosted regime by the ATLAS experiment, and Section~\ref{sec:HWW} presents results by both ATLAS and CMS in the \HWW decay channel.

\section{Boosted \Hbb Decays with ATLAS}\label{sec:Hbb}
% ATLAS Boosted H(bb)
% https://atlas.web.cern.ch/Atlas/GROUPS/PHYSICS/CONFNOTES/ATLAS-CONF-2021-010/

While unexplored areas where new physics discoveries could be hiding are numerous, one of the regions of phase space for the Higgs decay with a large sensitivity to new physics is the so-called boosted regime, where the Higgs boson has sizable transverse momentum.
This regime is difficult to explore due to the challenging nature of reconstructing the Higgs boson decay products.
In this analysis, the Higgs boson candidate is reconstructed as a single large-radius jet containing two b-tagged track-jets.
A second jet is required to ensure a di-jet topology.
Either the leading or sub-leading large-radius jet can be considered as the Higgs candidate.
The dominant background process for this measurement are multi-jet events, which are modeled by an analytic function.
The measurement is inclusive in production modes; Table~\ref{tab:Hbb} gives the relative fraction of the four main Higgs production modes present in the fiducial measurements carried out in this analysis.

\begin{table}[t]
       \caption[]{Signal acceptance times efficiency for the signal regions in the fiducial measurements~\cite{ATLAS_HBB_BOOSTED}.}
       \label{tab:Hbb}
       \vspace{0.4cm}
       \begin{center}
              \begin{tabular}{l c c}
              \toprule
              Process & $\ptH > 450~\GeV$ & $\ptH > 1~\TeV$\\
              \hline
              All & 0.25 & 0.18\\
              \hline
              \ggH & 0.26 & 0.22\\
              \VH & 0.27 & 0.19\\
              \VBF & 0.22 & 0.15\\
              \ttH & 0.20 & 0.16\\
              \bottomrule
              \end{tabular}
       \end{center}
\end{table}

Inclusive, fiducial, and differential measurements are performed via a binned maximum-likelihood fit to the Higgs candidate jet mass distribution.
The event yields measured in this analysis correspond to Higgs boson production \xs values in the fiducial region of

\begin{equation*}
       \begin{array}{rcl}
              \sigma_H(\ptH > 450~\GeV) &=& 13  \pm 57  \text{ (stat.)} \pm 22  \text{ (syst.)} \pm 3   \text{ (theo.})~\fb,\\
              &&\\
              \sigma_H(\ptH > 650~\GeV) &=& 13  \pm 16  \text{ (stat.)} \pm 7   \text{ (syst.)} \pm 3   \text{ (theo.})~\fb,\\
              &&\\
              \sigma_H(\ptH > 1~\TeV)   &=& 3.4 \pm 3.9 \text{ (stat.)} \pm 1.0 \text{ (syst.)} \pm 0.8 \text{ (theo.})~\fb.\\
              &&\\
       \end{array}
\end{equation*}

These measurements are currently statistically dominated, and provide a first measurement in the phase space $\ptH > 1~\TeV$.

\section{\HWW Decays}\label{sec:HWW}

In the following, Section~\ref{subsec:ATLAS_HWW} presents the latest result from the ATLAS Collaboration, focusing on the \ggH and \VBF production modes, and Section~\ref{subsec:CMS_HWW} describes the CMS analysis measuring the \VH production mode.

\subsection{\ggH and \VBF ATLAS Measurements}\label{subsec:ATLAS_HWW}
% ATLAS H(WW)
% https://atlas.web.cern.ch/Atlas/GROUPS/PHYSICS/CONFNOTES/ATLAS-CONF-2021-014/
This analysis measures the \ggH and \VBF production modes of the Higgs boson, and its subsequent decay into a pair of $W$ bosons, with a final state consisting of an electron, a muon, and missing transverse momentum.
Events are required to contain an opposite-sign muon-electron pair and missing transverse momentum.
Either a single-lepton or electron-muon trigger is required to have fired.
Most background processes are simulated with Monte Carlo methods, however the contribution from non-prompt or fake leptons is derived from data.
Compared to the previous Run 2 results from ATLAS, several improvements to the analysis have been incorporated in addition to the increase in data statistics from 36~\ifb to 139~\ifb, most notably a measurement of the \ggH production mode in the final state with two or more reconstructed jets and measurements of \xss in kinematic fiducial regions defined in the STXS framework.
A maximum likelihood fit to the discriminating variable distributions (either the dilepton transverse mass for the \ggH production mode or the output of a deep neural network for the \VBF production mode) is performed in order to extract the parameters of interest.
STXS measurements are provided in a total of 11 categories, as shown in Figure~\ref{fig:ATLAS_HWW_STXS}. 

\fig[0.7]{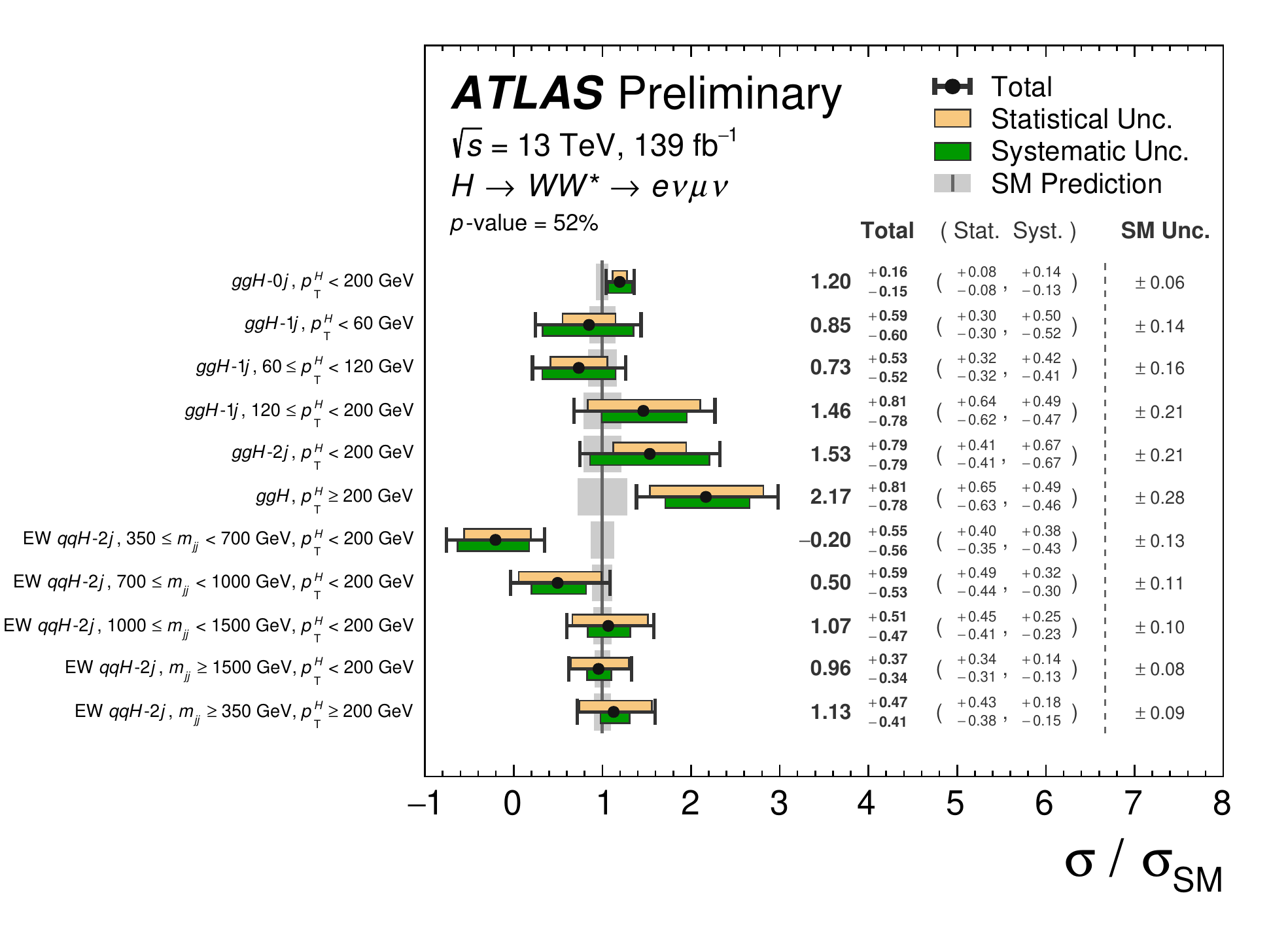}
  {Best-fit values and uncertainties for the \xss measured in each of the STXS categories, normalised to the corresponding SM predictions. The black error bars, green boxes and tan boxes show the total, systematic, and statistical uncertainties in the measurements, respectively. The grey band represents the theory uncertainty on the signal production corresponding to the STXS category~\cite{ATLAS_HWW}.}
  {fig:ATLAS_HWW_STXS}

\subsection{$V\HWW$ CMS Measurements}\label{subsec:CMS_HWW}
% CMS VH(WW)
% http://cms-results.web.cern.ch/cms-results/public-results/preliminary-results/HIG-19-017/
This result presents measurements of \VH production in the \HWW decay channel, where at least one $W$ boson decays leptonically.
The \VH production mode provides a direct probe of the Higgs boson coupling to vector bosons, and \VH kinematics are sensitive to the effects of new physics beyond the SM.
The analysis selects events with final states containing two to four leptons, and simulates the majority of the background processes with Monte Carlo methods, with the exception of the non-prompt or fake lepton backgrounds, which are derived from data.
A maximum likelihood fit to the discriminating variable distributions, either the transverse mass or the output of a boosted decision tree, is performed in order to obtain the final measurements.
The signal strength, defined as the ratio between the measured signal \xs and the SM expectation, provides insight on the compatibility between the measurements and the SM.
The measurements are carried out in 4 STXS bins, shown in Figure~\ref{fig:CMS_HWW_STXS}, where \ptV corresponds to the transverse momentum of the associated vector boson $V$.
The $\ptV < 150~\GeV$ phase space has been measured in this channel for the first time.

\fig[0.7]{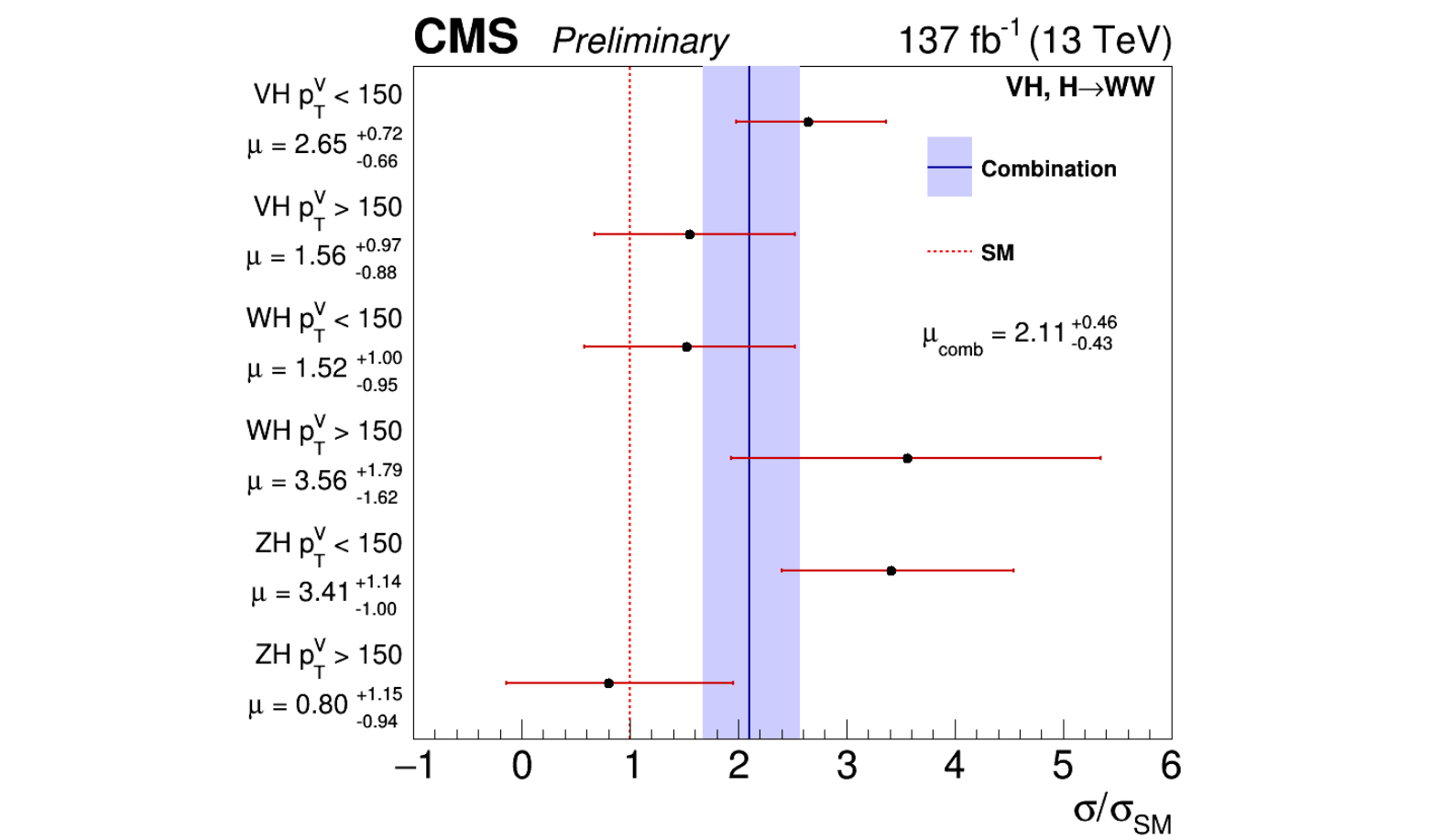}
  {Comparison between the combined signal strength, determined by applying a single signal strength to both \ptV bins and all \VH production modes in the STXS fit, and the signal strengths for each \ptV bin and production mode ($WH$, $ZH$, and \VH inclusive)~\cite{CMS_HWW}.}
  {fig:CMS_HWW_STXS}

\section{Conclusion}
Tremendous progress has been made in measuring properties and \xss of the Higgs boson at the LHC since its discovery 9 years ago.
The field is making big improvements in the precision of a large number of Higgs measurements, both from using the full Run 2 dataset, and from continually improving analysis methods.
The ATLAS and CMS Collaborations are still only at the early stages of exploring the Higgs sector; with the start of Run 3 approaching, new data and more analysis improvements will provide unprecedented levels of precision, and potentially hints of new physics beyond the SM.

\section*{Acknowledgments}

The author would like to thank the organizers for making the 55$^\text{th}$ Rencontres de Moriond a success, and acknowledges the support of the Natural Sciences and Engineering Research Council of Canada (NSERC).

\printbibliography

\noindent Copyright 2021 CERN for the benefit of the ATLAS and CMS Collaborations. Reproduction of this article or parts of it is allowed as specified in the CC-BY-4.0 license.

\end{document}